\documentclass{llncs}
\usepackage{multicol}
\usepackage{cite}
\usepackage{adjustbox}
\usepackage{makeidx}  
\usepackage{tabularx}
\usepackage{multirow}
\usepackage{graphicx}
\usepackage{booktabs} 
\usepackage{amsmath}  
\usepackage{color}    
\usepackage[utf8]{inputenc} 
\usepackage{caption}
\usepackage{etoolbox} 
\makeatletter
\patchcmd{\ps@headings}{\rlap{\thepage}}{}{}{}
\patchcmd{\ps@headings}{\llap{\thepage}}{}{}{}
\makeatother
\begin{document}
%
%

%
\mainmatter              
\title{Shape-Sensitive Loss for Catheter and Guidewire Segmentation}
\titlerunning{Hamiltonian Mechanics}  
%

%
\author{Chayun Kongtongvattana\inst{1} \and Baoru Huang\inst{2} \and Jingxuan Kang\inst{1} \and Hoan Nguyen\inst{3,4} \and Olajide Olufemi\inst{5} \and Anh Nguyen\inst{1}}
\authorrunning{Kongtongvattana et al.} 
%
\tocauthor{Ivar Ekeland, Roger Temam, Jeffrey Dean, David Grove,
Craig Chambers, Kim B. Bruce, and Elisa Bertino}

\institute{
Department of Computer Science, University of Liverpool, UK \and Imperial College London, UK \and 
University of Information Technology, HCMC, Vietnam \and
Vietnam National University, HCMC, Vietnam \and 
Alder Hey Children's Hospital, Liverpool, UK
}

\maketitle              

\begin{abstract}
We introduce a shape-sensitive loss function for catheter and guidewire segmentation and utilize it in a vision transformer network to establish a new state-of-the-art result on a large-scale X-ray images dataset. We transform network-derived predictions and their corresponding ground truths into signed distance maps, thereby enabling any networks to concentrate on the essential boundaries rather than merely the overall contours. These SDMs are subjected to the vision transformer, efficiently producing high-dimensional feature vectors encapsulating critical image attributes. By computing the cosine similarity between these feature vectors, we gain a nuanced understanding of image similarity that goes beyond the limitations of traditional overlap-based measures. The advantages of our approach are manifold, ranging from scale and translation invariance to superior detection of subtle differences, thus ensuring precise localization and delineation of the medical instruments within the images. Comprehensive quantitative and qualitative analyses substantiate the significant enhancement in performance over existing baselines, demonstrating the promise held by our new shape-sensitive loss function for improving catheter and guidewire segmentation.
\end{abstract}
\section{Introduction}\label{sec1}
Endovascular interventions have drastically changed cardiovascular surgery, bringing forth benefits like minimized trauma and faster recovery. However, they also present potential hazards such as damage to the vessel wall~\cite{rafii-tari2014, MedTech2018,cathsim}. Accurate segmentation of catheters and guidewires within X-ray images is pivotal for reducing these risks.

\begin{figure}[h]
\centering
\includegraphics[width=0.9\linewidth]{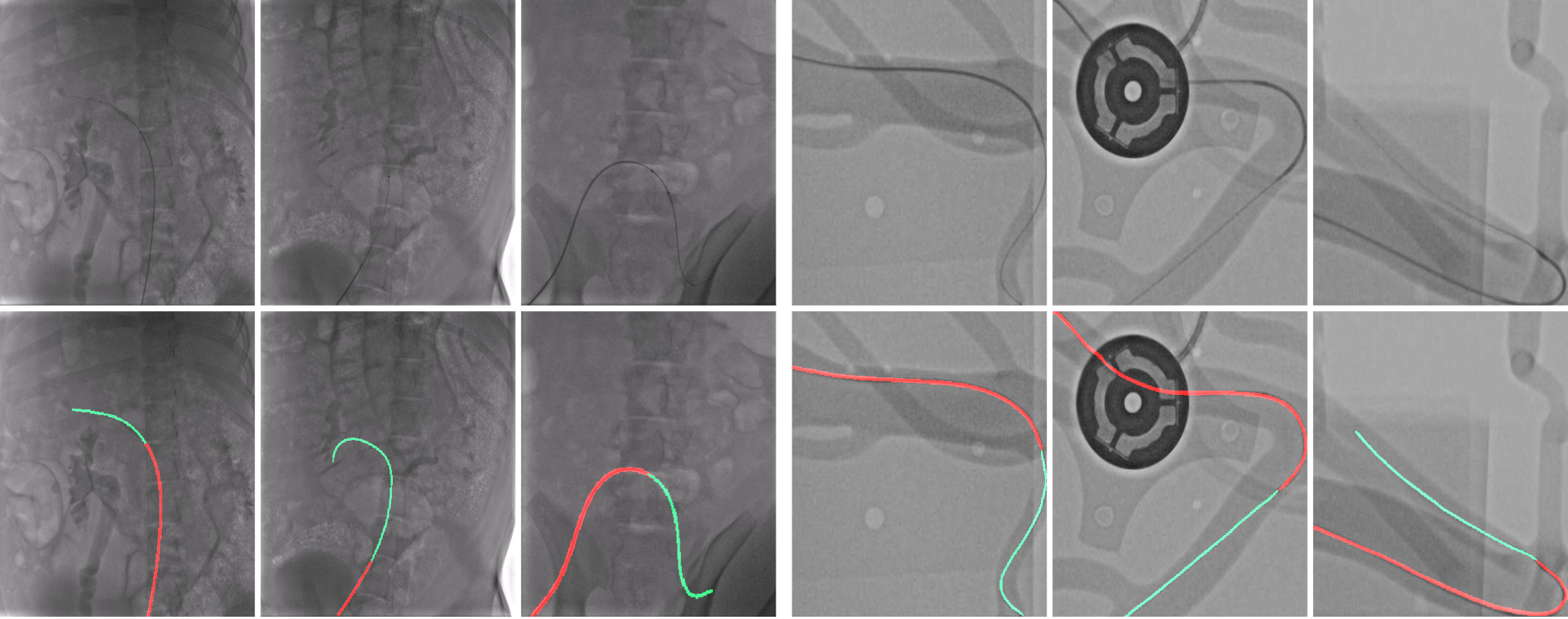}
\caption{Catheter and guidewrite segmentation in X-ray images. \textbf{First row:} The input X-ray images. \textbf{Second row:} The segmentation results. Red color denotes the catheter, green color denotes the guidewire.}
\label{fig:result_real}
\end{figure}

The merging of computer vision and machine learning in the medical domain has spurred advancements in addressing challenges associated with endovascular interventions~\cite{thakur2009, abdelaziz2019, bian2013, chi2020, zhao2019}. Especially, the role of deep learning has proven significant in refining the precision of surgical procedures and enhancing patient safety~\cite{benavente2019,sensing_det}. Despite these strides, the segmentation of intricate structures of catheters and guidewires in X-ray images remains a hurdle. Traditional loss functions often fall short in capturing the global spatial relationships crucial for this task~\cite{oshea2015introduction,depth_mi, Jadon_2020}. While research has employed convolutional neural networks (CNNs) with these loss functions to deliver promising results~\cite{ShuaiGuo2019,lw_tmi, gherardini2020catheter, s21082615}, the challenge persists.

Our study presents a novel shape-sensitive loss function that uses the Vision Transformer (ViT) to gauge the similarity between signed distance maps (SDMs)~\cite{dosovitskiy2021image, 9433775}. This approach aims to offer superior context awareness and structure sensitivity, leading to enhanced segmentation.

The structure of this paper comprises a literature review, a detailed explanation of our proposed loss function, presentation of experimental results, and a concluding section discussing potential future directions.

Outlined below are the foundational contributions of this work:
\begin{itemize}
\item Introduction of a shape-sensitive loss function that merges spatial distance insights with the feature extraction prowess of Vision Transformer.
\item A unique technique for converting network outputs into SDMs, preserving structural intricacies.
\item A balanced approach, merging our shape-sensitive loss with the traditional Dice loss to ensure holistic segmentation performance.
\end{itemize}

\section{RELATED WORKS}\label{sec2}


\subsection{Catheter Segmentation}

Catheter segmentation has gained momentum with the introduction of deep learning frameworks~\cite{Ambrosini_miccai,nguyen2020endtoend}. The FW-Net utilizes an end-to-end approach with an encoder-decoder structure, optical flow extraction, and a unique flow-guided warping function to ensure temporal continuity in imaging sequences~\cite{nguyen2020endtoend}. Another method employs deep convolutional neural networks for segmenting catheters and guidewires in 2D X-ray fluoroscopic sequences, using previous image contexts for enhanced accuracy and achieving a notable median centerline distance error of 0.2 mm~\cite{ambrosini2017fully}. A transformative approach incorporates Convolutional Neural Networks (CNNs) with transfer learning, exploiting synthetic fluoroscopic images to develop a streamlined segmentation model requiring minimal manually annotated data, significantly reducing testing time while remaining adaptable to higher input resolutions~\cite{gherardini2020catheter}. These advancements underscore the continuous evolution and adaptability of deep learning methodologies in catheter and guidewire segmentation tasks.

\vspace{-1em}
\begin{table}[h]
\caption{Types of Semantic Segmentation Loss functions}\label{table:simplified_loss_functions}
\begin{center}
\begin{tabular}{@{}ll@{}}
\toprule
\multicolumn{2}{c}{\textbf{Distribution-Based Loss}} \\
\midrule
Binary Cross-Entropy~\cite{ma2004automated} & $-\sum y \log(p) + (1-y) \log(1-p)$ \\
Weighted Cross-Entropy~\cite{pihur2007weighted} & $-\sum w_y y \log(p)$ \\
Balanced Cross-Entropy~\cite{xie2015holistically} & $-\beta \sum y \log(p) - (1-\beta) \sum (1-y) \log(1-p)$ \\
Focal~\cite{lin2017focal} & $-\sum (1-p)^{\gamma} y \log(p)$ \\
\midrule
\multicolumn{2}{c}{\textbf{Region-Based Loss}} \\
\midrule  
Dice~\cite{sudre2017generalised} & $1 - \frac{2 \sum (y \cap p)}{\sum y + \sum p}$ \\
Tversky~\cite{salehi2017tversky} & $\frac{\sum (y \cap p)}{\sum (y \cap p) + \alpha \sum (y - p) + \beta \sum (p - y)}$ \\
Focal Tversky~\cite{abraham2019novel} & $(1 - Tversky)^{\gamma}$ \\
\midrule
\multicolumn{2}{c}{\textbf{Compound Loss}} \\
\midrule  
Combo~\cite{taghanaki2019combo} & $CL(y,\hat{y}) = \alpha L_{m-bce} - (1-\alpha) DL(y,\hat{y})$ \\
ELL~\cite{wong20183d} & $L_{ELL} = \alpha L_{Dice} + \beta L_{CE}$ \\
\midrule
\multicolumn{2}{c}{\textbf{Boundary-Based Loss}} \\
\midrule  
HD~\cite{karimi2019reducing} & $L_{HD_{DT}} = \frac{1}{N} \sum_{i=1}^{N} [(s_i - g_i) \cdot (d_{Gi}^2 - d_{Si}^2)]$ \\
InverseForm~\cite{borse2021inverseform} & $L_{\text{if}}(b_{\text{pred}}, b_{\text{gt}}) = \sum_{j=1}^{N} d_{\text{if}}(b_{\text{pred,j}}, b_{\text{gt,j}})$ \\
Shape-sensitive \textcolor{blue}{(ours)} & Equation \ref{eq:shape_sensitive_loss} \\
\bottomrule
\end{tabular}
\end{center}
\end{table}
\vspace{-1em}

\subsection{Loss Function for Medical Segmentation}
Loss functions in deep learning for image segmentation tasks are pivotal for determining the quality of medical image segmentations (Table~\ref{table:simplified_loss_functions}). These functions can be broadly categorized into four primary types~\cite{Jadon_2020}: distribution-based, region-based, boundary-based, and compounded loss. Distribution-based losses, such as Binary Cross-Entropy~\cite{ma2004automated}, Focal loss~\cite{lin2017focal}, and Weighted Cross-Entropy~\cite{pihur2007weighted}, measure the dissimilarity between predicted and true probability distributions. While effective in modeling these distributions, they might lack spatial coherence and can struggle with class imbalances. On the other hand, region-based losses like Dice loss~\cite{sudre2017generalised}, Tversky Loss~\cite{salehi2017tversky}, and Focal Tversky loss~\cite{abraham2019novel} excel in scenarios with class imbalances by focusing on the overlap between predicted and actual segments, but they may miss finer details. Boundary-based losses, such as Shape-aware loss~\cite{hayder2016shapeaware} and Hausdorff Distance loss~\cite{karimi2019reducing}, prioritize boundary accuracy but can be sensitive to minor perturbations. Meanwhile, compounded losses like Combo loss~\cite{taghanaki2019combo} and Exponential Logarithmic Loss~\cite{wong20183d} provide a comprehensive approach by merging features from different loss types, though they can potentially increase training complexity. 

The pursuit of accuracy, especially for intricate structures, led to the inception of shape-aware loss functions, which factor in the spatial relationships and distances of pixels from target boundaries. The InverseForm loss function~\cite{borse2021inverseform} is a noteworthy example. It integrates an Inverse Transformation Network into the loss calculation to produce a Transformation Matrix, ensuring alignment between predicted segmentation and target boundaries. Nonetheless, capturing high-level features that provide global context can be challenging for some methods. Drawing inspiration from such integrative approaches, our proposal incorporates Vision Transformers (ViTs) within the loss function. With their attention mechanism, ViTs capture global contextual features from images, paving the way for a more accurate representation of complex structures, such as catheters and guidewires in X-ray images.

\section{Shape-Sensitive Loss for Catheter and Guidewire Segmentation}\label{sec3}

In this section, we introduce a novel approach enhancing catheter and guidewire segmentation in X-ray images. Leveraging SDM and ViT foundations, our method integrates key modifications to improve accuracy, offering a fresh perspective in precise medical imaging.

\subsection{Preliminaries: Signed Distance Map}\label{subsec:SDM}

\begin{figure}[t]
    \centering
    \includegraphics[width=0.6\linewidth]{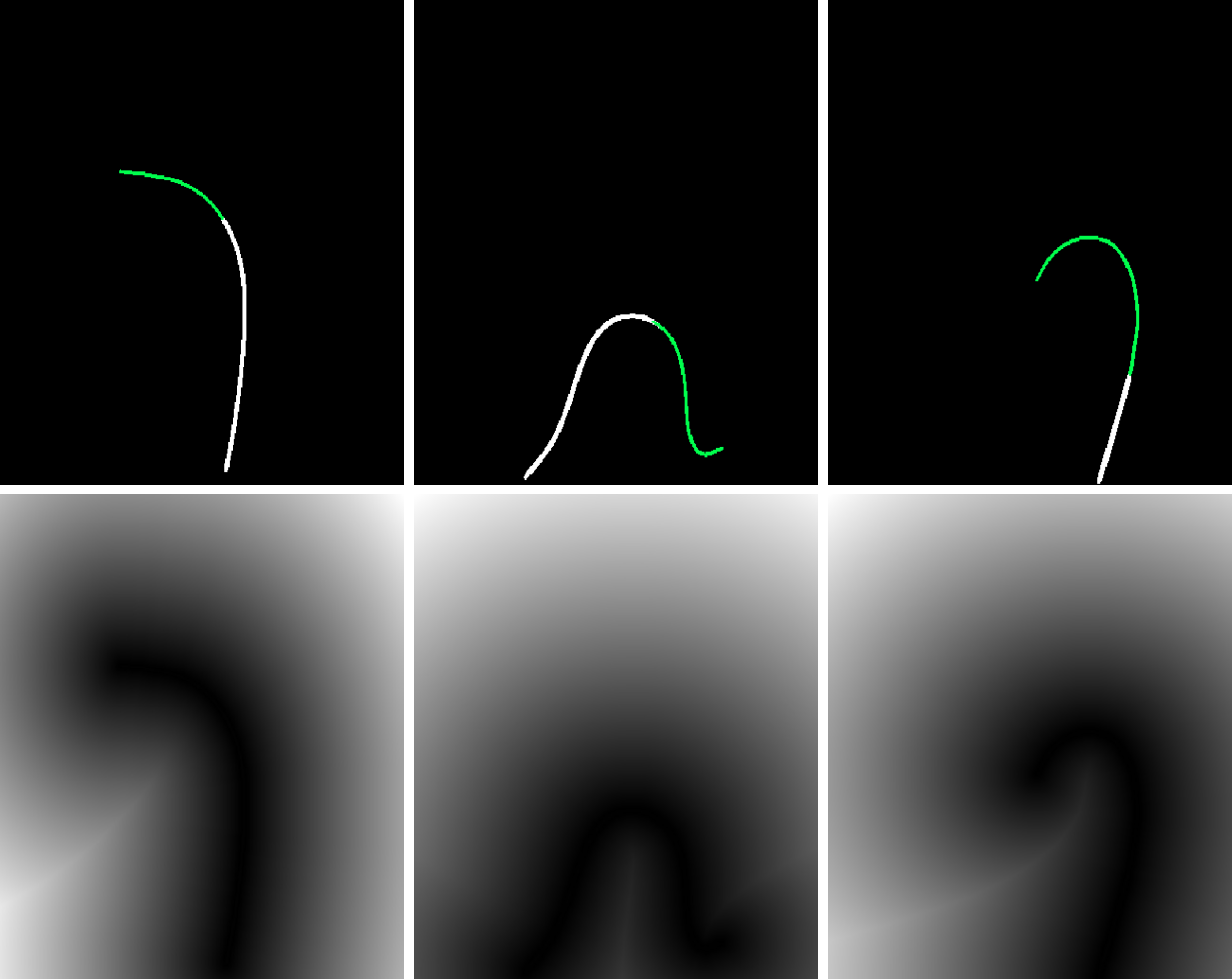}
    \caption{Illustration of the process to create the Signed Distance Maps. \textbf{Top Row:} Original groundtruth images. \textbf{Bottom Row:} Signed Distance Maps, calculated based on the contours, overlaid by its contour images}
    \label{fig:SDM}
\end{figure}

The Signed Distance Map (SDM) is reshaping image segmentation by mapping each pixel's distance to the image's contour, signifying its relation to the boundary. This method enhances the clarity of segmentation, especially for complex structures. SDM's emphasis on boundary localization is crucial in medical imaging, like catheter and guidewire segmentation. By using SDM, models become more sensitive to boundaries, prioritizing localization and improving segmentation accuracy, resulting in fewer errors in X-ray image segmentation. Fig.~\ref{fig:SDM} shows an example of SDM.

\textbf{Initial Transformation to SDM:} Both the predicted output and its corresponding label undergo an initial transformation to fit the SDM format, setting the stage for further operations tailored for this representation. Let $N(x,y)$ be the network output, where $(x,y)$ are pixel coordinates.

Transformation to Contour: The contour representation, $C(x,y)$, is derived by thresholding the network output at a value $T:$

\begin{equation}
C(x, y) =
\begin{cases}
1 & \text{if } N(x, y) > T \\
0 & \text{otherwise}
\end{cases}
\label{eq:my_equation}
\end{equation}

Transformation to SDM: Define $d((x,y),(i,j))$ as the Euclidean distance between any point $(x,y)$ and its nearest boundary point $(i,j).$
\begin{equation}
\text{SDM}(x, y) = 
\begin{cases} 
0, & \text{if } C(x, y) = 1 \\
\min_{(i, j)} d((x, y), (i, j)), & \text{if } C(x, y) \neq 1 
\end{cases}
\end{equation}

The computation of $d((x,y),(i,j))$ often employs methods like the Fast Marching Method or the Distance Transform algorithm. The SDM produces a continuous spectrum of values, with positive distances outside the object and negative distances inside. Leveraging SDM in the loss function allows the network to achieve precise segmentation of intricate structures in X-ray images.

\subsection{Feature Extraction using Vision Transformer}\label{subsec:ViT}

The Vision Transformer (ViT) excels in extracting high-level features, pivotal for understanding the intricate patterns in Signed Distance Maps (SDMs). Our approach, while building upon ViT's established use with SDM, introduces key modifications to enhance feature extraction. These modifications, including optimized patch sizes for high-resolution imaging and an adaptive attention mechanism targeting critical anatomical features, are specifically tailored to address the variability in medical data and the need for precise segmentation. This integration not only demonstrates ViT's potential in specialized tasks, but also distinctly sets apart our method, as shown in Fig.~\ref{fig:Architecture}.

For feature extraction, the last layer of the ViT is bypassed, allowing it to produce high-dimensional vectors that capture detailed SDM patterns—a critical component for precise segmentation. Opting for the ViT-B/384 configuration, designed for high-resolution images, is ideal. Using a 384-patch size, it captures fine-grained details, making it apt for high-resolution SDM analysis. From this, features $\alpha$ for predicted and $\beta$ for true SDMs are determined, forming the core for loss computation.

\begin{equation}
\alpha, \beta = \text{ViT}(\text{SDM}_{\text{Output}}, \text{SDM}_{\text{Label}})
\end{equation}

In the equation above, $\text{ViT}(\text{SDM}_{\text{Output}}, \text{SDM}_{\text{Label}})$ demonstrates the feature extraction process by passing SDMs through the ViT. By utilizing ViT, segmentation is further enhanced, resulting in a thorough of SDM structures.

\subsection{Shape-Sensitive Loss Computation}\label{subsec:Loss}

\begin{figure}[t]
\centering
\includegraphics[width=1\textwidth]{img/SSLoss_Diagram_V4_Good.pdf}
\caption{An overview of our framework.}
\label{fig:Architecture}
\end{figure}

Accurate segmentation heavily hinges on an effective loss computation mechanism to steer the model's learning, ensuring the precise identification of segmentation boundaries within SDMs. The employed loss computation in this work is fundamentally shape-sensitive, an approach meticulously tailored to heighten the model’s sensitivity to the geometric details within the SDMs. This heightened sensitivity is achieved through the use of cosine similarity between the high-level features extracted from both the predicted and true SDMs.

The premise behind employing cosine similarity lies in its capacity to consider the angle between the feature vectors, offering a nuanced and detailed insight into their alignment. This perspective enables a more delicate evaluation of the segmentation process, capturing the geometric intricacies within the high-dimensional feature space derived from the SDMs.

\begin{equation}
\text{CosSim } \left(\alpha , \beta\right) = \frac{\left(\alpha \cdot \beta\right)}{\left( | \alpha | \times | \beta | \right)}
\end{equation}

The shape-sensitive loss, denoted as $L_{\text{SS}}$, is thus defined as the deviation from perfect alignment, represented mathematically as:

\begin{equation}
\begin{split}
\mathcal{L}_{\text{SS}}(\text{SDM}_{\text{Output}}, \text{SDM}_{\text{Label}})
& = \\1 - \operatorname{CosSim}(\text{ViT}(\text{SDM}_{\text{Output}}),
& \quad \text{ViT}(\text{SDM}_{\text{Label}}))
\end{split}
\label{eq:shape_sensitive_loss}
\end{equation}

Further enriching the loss computation, the total loss, $\mathcal{L}_{\text{total}}$, amalgamates the shape-sensitive loss with the Dice loss, a well-regarded loss function for segmentation tasks. This composite measure is meticulously balanced for optimal segmentation results, ensuring the model maintains a comprehensive focus, attending to shape sensitivity and other pivotal aspects of segmentation.

\begin{equation}
\mathcal{L}_{\text{total}} = \gamma \mathcal{L}_{\text{Dice}} + \delta \mathcal{L}_{\text{SS}}(\text{SDM}_{\text{Output}}, \text{SDM}_{\text{Label}})
\end{equation}

This structured loss computation strategy, by intertwining geometric sensitivity with established loss functions, furnishes the model with a robust and multifaceted learning signal, underpinning the attainment of superior segmentation performance in processing SDMs.

\section{Experimental Results}\label{sec4}

\subsection{Experiment Setup}

\textbf{Dataset:} We assessed our proposed loss function using our newly collected dataset. This dataset includes 5,086 real animal X-ray images and 18,791 phantom X-ray images, both paired with ground truth annotations. While real animal images are inherently 512 × 512 pixels, phantom images were resized from 1024 × 1024 to 512 × 512, ensuring labels remained accurate. The dataset was divided with a 70-30 split for training and testing, respectively.

\textbf{Evaluation metrics:} Our semantic segmentation performance was gauged using several standard metrics:

\begin{itemize}
\item \textbf{Dice Coefficient:} Measures overlap between prediction and ground truth.
\item \textbf{Jaccard Similarity (IoU):} Calculates the ratio of intersected region to the combined predicted and ground truth areas.
\item \textbf{Mean Intersection over Union (mIoU):} This is an average IoU across all classes, predominantly for multi-class segmentation.
\item \textbf{Accuracy:} Determines the proportion of correctly identified pixels.
\end{itemize}

We integrate our method into different segmentation backbones, including U-Net~\cite{U-Net_paper},  U-Net++~\cite{U-Net++_paper}, U-Net3+~\cite{U-Net3+_paper}, TransU-net~\cite{TransU-Net_paper}, and SwinU-Net~\cite{SwinU-Net_paper}.

\subsection{Results on Real Animal X-Ray Images}


As shown in Table \ref{table:comparison}, all segmentation networks profited from our method. TransU-Net exhibited a boost in the Dice coefficient from 54.52\% to 57.16\%, a gain of 2.64 percentage points. Similar growth was observed in Jaccard Similarity, mIoU, and Accuracy. Expanding on this, the U-Net, a foundational segmentation architecture, after embedding our loss function, showed improvements in all four metrics, with the Dice coefficient rising by 1.75 percentage points. The other models, including U-Net++, U-Net3+, and SwinU-Net, mirrored this enhancement trend. Worth highlighting is that the TransU-Net, when synergized with our loss function, outperformed other architectures in all metrics. 

\vspace{-1em}
\begin{table}[htbp]
\centering
\caption{Comparing results for real animal X-ray images}
\label{table:comparison}
\begin{tabular*}{\columnwidth}{@{\extracolsep{\fill}}lcccc}
\toprule
\multicolumn{1}{c}{\textbf{Network}} & \textbf{Dice} & \textbf{Jaccard} & \textbf{mIoU} & \textbf{Accuracy} \\ 
\midrule
TransU-net & 54.52 & 43.87 & 61.15 & 78.25 \\ 
TransU-Net \textbf{\textcolor{blue}{w/ours}} & 57.16 & 46.04 & 62.66 & 79.57 \\ 
U-Net & 46.20 & 36.62 & 54.14 & 71.57 \\ 
U-Net \textbf{\textcolor{blue}{w/ours}} & 47.95 & 38.79 & 55.05 & 72.89 \\ 
U-Net++ & 48.77 & 39.22 & 56.17 & 72.30 \\ 
U-Net++ \textbf{\textcolor{blue}{w/ours}} & 50.35 & 41.39 & 57.08 & 73.62 \\ 
U-Net3+ & 49.24 & 40.12 & 56.37 & 73.14 \\ 
U-Net3+ \textbf{\textcolor{blue}{w/ours}} & 51.37 & 42.29 & 57.28 & 74.46 \\ 
SwinU-Net & 52.74 & 42.14 & 57.48 & 77.67 \\ 
SwinU-Net \textbf{\textcolor{blue}{w/ours}} & 54.71 & 44.58 & 59.04 & 79.13 \\ 
\bottomrule
\end{tabular*}
\end{table}
\vspace{-1em}

\subsection{Results on Phantom X-ray Images}

In our expanded study focusing on phantom X-ray images, the capabilities of our proposed method were prominently showcased, emphasizing its prowess in segmentation tasks (Table~\ref{table:comparison_phantom}). When we integrated our method with the TransU-Net model, we witnessed a significant improvement in the Dice coefficient. The metrics rose from a base of 40.83\% to an enhanced 43.71\%, validating our method's effectiveness in enhancing the segmentation accuracy of intricate medical images. A closer inspection of Figure \ref{fig:Param_phantom} brings forth another salient feature of our approach. Across different architectures, there was a consistent pattern - the parameter count remained stable. Interestingly, this consistency held even when our innovative loss function was introduced. This is a crucial observation as it suggests that our method not only bolsters segmentation performance but also maintains it without imposing any supplementary computational burdens. 


\begin{table}[htbp]
\centering
\caption{Comparing results for phantom X-ray images}
\label{table:comparison_phantom}
\begin{tabular*}{\columnwidth}{@{\extracolsep{\fill}}lcccc}
\toprule
\multicolumn{1}{c}{\textbf{Network}} & \textbf{Dice} & \textbf{Jaccard} & \textbf{mIoU} & \textbf{Accuracy} \\ 
\midrule
TransU-net & 40.83 & 33.94 & 46.01 & 66.18 \\ 
TransU-Net \textbf{\textcolor{blue}{w/ours}} & 43.71 & 36.65 & 47.21 & 67.58 \\ 
U-Net & 34.51 & 27.62 & 43.14 & 60.57 \\ 
U-Net \textbf{\textcolor{blue}{w/ours}} & 37.13 & 28.56 & 45.83 & 62.29 \\ 
U-Net++ & 35.24 & 27.43 & 44.72 & 61.76 \\ 
U-Net++ \textbf{\textcolor{blue}{w/ours}} & 37.25 & 28.94 & 45.97 & 63.12 \\ 
U-Net3+ & 35.53 & 27.87 & 44.89 & 61.88 \\ 
U-Net3+ \textbf{\textcolor{blue}{w/ours}} & 38.07 & 29.54 & 46.21 & 63.48 \\ 
SwinU-Net & 39.44 & 31.27 & 45.29 & 65.37 \\ 
SwinU-Net \textbf{\textcolor{blue}{w/ours}} & 41.17 & 33.79 & 46.23 & 66.88 \\ 
\bottomrule
\end{tabular*}
\end{table}


\subsection{Ablation Study}

\begin{figure}[htbp]
    \centering
    \begin{minipage}[t]{0.48\textwidth}
        \vspace{0pt}
        \centering
        \setlength\tabcolsep{5pt} 
        \renewcommand{\arraystretch}{1.0} 
        \captionof{table}{Effects of Blending Parameters: \(\gamma\) for Dice Loss and \(\delta\) for Our Proposed Shape-Sensitive Loss on Various Network Architectures.}\label{table:coefficients}
        \begin{tabularx}{\linewidth}{@{}cccc@{}}
            \toprule
            \multirow{2}{*}{\textbf{Network}} & \multicolumn{2}{c}{\textbf{Coefficient}} & \multirow{2}{*}{\textbf{Dice (\%)}} \\
            & \textbf{$\gamma$} & \textbf{$\delta$} & \\ 
            \midrule
            \multirow{5}{*}{TransU-Net} & 0.1 & 0.9 & 56.68 \\  
            & 0.3 & 0.7 & 57.03 \\  
            & 0.5 & 0.5 & 57.16 \\  
            & 0.7 & 0.3 & 56.91 \\  
            & 0.9 & 0.1 & 56.89 \\ 
            \midrule
            \multirow{5}{*}{SwinU-Net} & 0.1 & 0.9 & 54.15 \\  
            & 0.3 & 0.7 & 54.23 \\  
            & 0.5 & 0.5 & 54.71 \\  
            & 0.7 & 0.3 & 54.55 \\  
            & 0.9 & 0.1 & 54.50 \\ 
            \bottomrule
        \end{tabularx}
    \end{minipage}
    \hfill
    \begin{minipage}[t]{0.48\linewidth}
        \vspace{0pt}
        \centering
        \includegraphics[width=\linewidth]{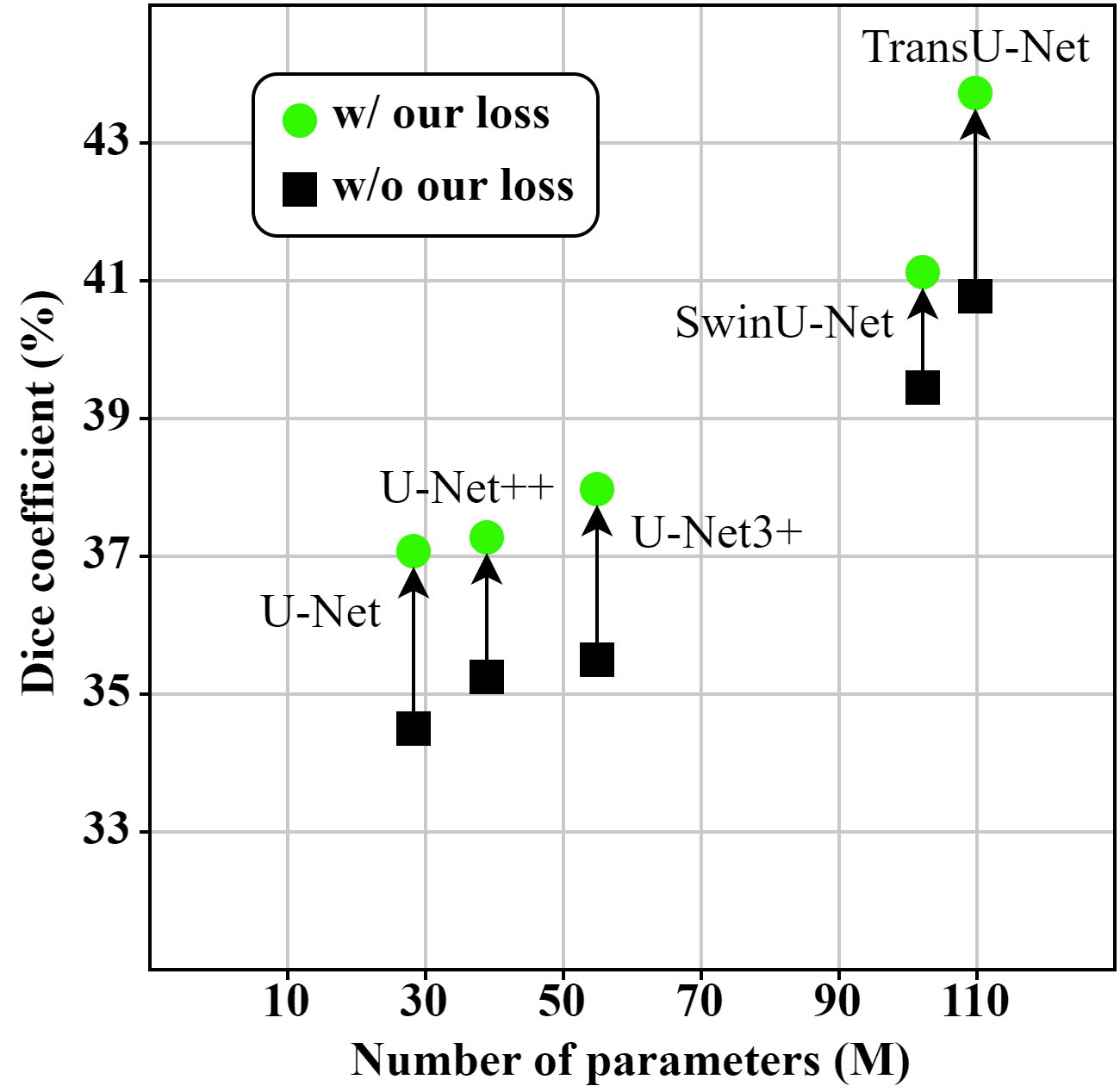}
        \caption{Comparison of Dice coefficient and number of parameters of different networks on phantom X-ray images.}
        \label{fig:Param_phantom}
    \end{minipage}
\end{figure}

In our exhaustive evaluation using real-animal X-ray image datasets, several critical findings became evident. Analyzing the data in Table \ref{table:Distance}, it is clear that the Cosine similarity consistently outperformed other traditional distance measurements when it came to evaluating feature embeddings. Further insights from Table \ref{table:Combination} revealed that integrating the Dice loss with our uniquely devised, shape-sensitive loss function culminated in the most robust segmentation results. Diving deeper, Table \ref{table:coefficients} offers a nuanced look into the impact of blending parameters across different network architectures. Notably, the TransU-net framework, when meticulously calibrated with blending coefficients of 0.5 for both ${\gamma}$ and ${\delta}$, stood out, delivering unparalleled performance. Summarizing our extensive assessments, the pinnacle was a Dice coefficient of 57.16\%, a testament to the potency and precision of our methodological choices and rigorous parameter adjustments in the demanding realm of medical image segmentation.

\newcolumntype{Y}{>{\centering\arraybackslash}X}
\vspace{-1em}
\begin{table}[htbp]
    \centering
    \small
    \renewcommand{\thempfootnote}{\arabic{mpfootnote}}
    \begin{minipage}[t]{0.48\textwidth}
        \centering
        \caption{Performance comparison of different loss functions.}\label{table:Combination}
        \begin{tabularx}{\linewidth}{@{}YY@{}}
            \toprule
            \textbf{Loss Function} & \textbf{Dice (\%)} \\
            \midrule
            CE\footnotemark[1] and Dice & 54.52 \\
            FT\footnotemark[2] and \textbf{\textcolor{blue}{ours}} & 55.78 \\
            Combo and \textbf{\textcolor{blue}{ours}} & 56.63 \\
            Dice and \textbf{\textcolor{blue}{ours}} & 57.16 \\
            \bottomrule
        \end{tabularx}
        \footnotetext[1]{CE = Cross-Entropy}
        \footnotetext[2]{FT = Focal Tversky}
    \end{minipage}
    \hfill
    \begin{minipage}[t]{0.48\linewidth}
        \centering
        \captionof{table}{Comparison between various distance measurements for loss value calculation.}
        \label{table:Distance}
        \begin{tabular}{@{}cc@{}}
            \toprule
            \textbf{Dist. Measurement} & \textbf{Dice (\%)} \\ 
            \midrule
            Cosine similarity & 57.16 \\  
            Euclidean distance & 56.78\\  
            Manhattan distance & 56.10\\ 
            Jaccard similarity & 56.72\\  
            Hamming distance & 54.03\\  
            \bottomrule
        \end{tabular}
    \end{minipage}
\end{table}
\vspace{-1em}

\section{Conclusion}\label{sec5}

In this study, we proposed a customized loss function for shape-sensitive segmentation using a pre-trained Vision Transformer (ViT) network. By converting the network prediction and ground truth segmentation maps into signed distance maps and extracting high-level features through the ViT, we were able to estimate feature matrices. The spatial distance between boundary maps was then evaluated using Cosine similarity, which measured the dissimilarity between the extracted high-level features. By combining our customized loss function with the Dice loss, we aimed to leverage shape-sensitive segmentation and capture finer details, ultimately improving the overall segmentation accuracy. Our approach demonstrated the effectiveness of utilizing ViT and signed distance map in the segmentation task, providing valuable insights into optimizing boundary map distances for accurate segmentation results. While our strategy underscored the merits of employing both ViT and signed distance map for segmentation, it's essential to acknowledge the relatively subdued accuracy in our prediction results. This highlights a pivotal area of potential enhancement, motivating us to delve deeper into refinements and optimizations to elevate the segmentation accuracy in future endeavours.

%
%

\end{document}